\documentclass[11pt,a4paper]{article}
\usepackage{jcappub}
\usepackage{euscript}
\usepackage{amsfonts,latexsym}

\newcommand{\be}[1]{\begin{equation}\label{#1}}
\newcommand{\ee}{\end{equation}}
\newcommand{\ba}[1]{\begin{eqnarray}\label{#1}}
\newcommand{\ea}{\end{eqnarray}}
\newcommand{\rf}[1]{(\ref{#1})}
\newcommand{\nn}{\nonumber}

\begin{document}

\title{Hubble flows and gravitational potentials in observable Universe}

\author{Maxim Eingorn} \author{and Alexander Zhuk}

\affiliation{Astronomical Observatory, Odessa National University,\\ Street Dvoryanskaya 2, Odessa 65082, Ukraine}

\emailAdd{maxim.eingorn@gmail.com} \emailAdd{ai.zhuk2@gmail.com}

\abstract{In this paper, we consider the Universe deep inside of the cell of uniformity. At these scales, the Universe is filled with inhomogeneously distributed
discrete structures (galaxies, groups and clusters of galaxies), which disturb the background Friedmann model. We propose mathematical models with conformally flat,
hyperbolic and spherical spaces. For these models, we obtain the gravitational potential for an arbitrary number of randomly distributed inhomogeneities. In the cases of
flat and hyperbolic spaces, the potential is finite at any point, including spatial infinity, and valid for an arbitrary number of gravitating sources. For both of these
models, we investigate the motion of test masses (e.g., dwarf galaxies) in the vicinity of one of the inhomogeneities. We show that there is a distance from the
inhomogeneity, at which the cosmological expansion prevails over the gravitational attraction and where test masses form the Hubble flow. For our group of galaxies, it
happens at a few Mpc and the radius of the zero-acceleration sphere is of the order of 1 Mpc, which is very close to observations. Outside of this sphere, the dragging
effect of the gravitational attraction goes very fast to zero.}


\maketitle

\flushbottom


\section{\label{sec:1}Introduction}

\setcounter{equation}{0}

According to astronomical observations, there is no clear evidence of spatial homogeneity up to sizes  $\sim$ 150 Mpc \cite{Labini} (see also \cite{Wilt1,Wilt2}). On the
other hand, on the larger scales the Universe is homogeneous and isotropic, being well described by the $\Lambda$CDM model \cite{7WMAP} which is the modern realization
of the Friedmann model. It is well known that the Hubble flow, i.e. the linear velocity-distance relation for receding motion of galaxies due to the expansion of the
Universe, is one of the characteristic features of the Friedmann model. The receding galaxies form the Hubble flows. Therefore, it seems natural to observe the Hubble
flows on scales larger than 150 Mpc. On the smaller scales peculiar velocities of inhomogeneities (galaxies, group of galaxies, etc) may distort these flows. Using the
characteristic values of peculiar velocities
$v\sim 200\div 400 \, \mbox{km}/\mbox{sec}$ and the present value of the Hubble parameter $H_0\approx 70$ km/sec/Mpc \cite{7WMAP}, we can get a rough estimate of
distances at which the Hubble flow velocity is of the same value as peculiar velocities: $D=v/H_0\approx 3\div 6$ Mpc. Hence, the Hubble flows can be observed at
distances greater than this rough estimate. From this point, it looks reasonable that Edwin Hubble discovered his law in 1929 after observing galaxies on scales less
than 20-30 Mpc. However, recent observations \cite{Sand1,Kar2003,Sand2,Kar2008,Kar2012} indicate the presence of Hubble flows at distances less than a few Mpc (even less
than 3 Mpc (see, e.g., \cite{Kar2008})) from the center of our group of galaxies. It means that peculiar velocities are sufficiently small to dilute considerably the
Hubble flow.  In other words, the local Hubble flow is cold \cite{Sand3,Kar2008,Kar2012}. Obviously, these observations require rigorous theoretical substantiation. It
is clear that deep inside of the cell of uniformity peculiar velocities are not the only reason which can destroy the Hubble flow. Here, the gravitational interaction of
the inhomogeneities plays an important role, and the rough estimate made above does not indicate at which distances from the gravitating objects the Hubble flow
overcomes this attraction. To resolve this important problem, it was suggested in \cite{Chernin1,CherninUFN1,CherninUFN2}  that the cosmological constant/dark energy is
responsible for the local cold flows. This idea was also supported in the papers based on observations \cite{Kar2003,Sand2,Kar2008}. However, this suggestion was
criticized in the article \cite{LR}, where the authors argue that at present time the peculiar velocities are sufficiently strong to dilute the Hubble flows in the local
region. To prove it, they used the hydrodynamical approach.

As we have mentioned above, for each inhomogeneity there is a local region where the gravitational attraction prevails over the cosmological expansion. Obviously, the
characteristic size of this region depends on the mass of the inhomogeneity. We can introduce the notion of the zero-acceleration surface where a free falling body
(which has the zero angular momentum) will have the zero acceleration. Roughly speaking, test bodies undergo the local dynamics inside of this surface and follow the
global expansion of the Universe outside of it. One of the main aims of our paper is to provide a rigorous definition of this surface. To do it, we consider our Universe
deep inside of the cell of uniformity. At such scales and on late stages of evolution, the Universe consists of a set of discrete inhomogeneities (galaxies, groups and
clusters of galaxies). Hence, classical mechanics of discrete objects provides more adequate approach than hydrodynamics with its continuous flows. Obviously,
inhomogeneities perturb the Friedmann-Robertson-Walker (FRW) metrics. In the weak field limit (the gravitational field of the inhomogeneities is weak and their
velocities are small with respect to the speed of light) such perturbations are reduced to the gravitational potential of the considered system (see, e.g.,
\cite{Mukhanov-book,Rubakov-book}). So, the first problem is to find this potential. In our paper we obtain a master equation for the gravitational potential of the
system (an arbitrary number of randomly distributed inhomogeneities) and solve this equation for models with conformally flat, hyperbolic and spherical spaces. In the
cases of flat and hyperbolic spaces, the potential is finite at any point, including spatial infinity, and valid for an arbitrary number of gravitating sources. In the
case of the flat space, to get such result, we need to cut off the gravitational potentials of each inhomogeneity at some concrete distances $r_0$ which we call the
radius of local gravity. The gravitational attraction of the inhomogeneities is absent outside of this radius. The cutoff is smooth, i.e. the potential is smoothly
matched at $r_0$. Such procedure looks a little bit artificial, but the hyperbolic case is free of this procedure. It is worth noting that the Schwarzschild-de Sitter
solution is valid only in the case of one gravitating mass and its potential is divergent at spatial infinity. Another serious drawback of this solution from the
cosmological point of view is that it does not take into account the average energy density of the matter in the Universe, which forms 27\% of the total energy density
at the present stage and depends on time. Our models are free from all of these defects.

Having the gravitational potential of an arbitrary system of inhomogeneities, we can investigate the motion of test masses (e.g., dwarf galaxies) in this field and the
formation of the Hubble flows by these masses. This approach is very convenient for numerical simulation. This is one of the main results of our paper. Additionally, for
both of our models, we define analytically the radii of the surfaces of the zero acceleration (the zero-acceleration radii).
For our group of galaxies,
they are of the order of 1 Mpc, which is very close to the observations \cite{Kar2012}. Outside of these surfaces, the dragging effect of the gravitational attraction
goes very fast to zero. The presence of the cosmological constant is not absolutely crucial for the Hubble flows. The reason for the Hubble flow is the global
cosmological expansion of the Universe. However, the cosmological constant provides the accelerating motion of the Hubble flows. Moreover, in the presence of the
cosmological constant, the Hubble flow is less diffused by peculiar velocities than in its absence (see, e.g., \cite{Chernin2004}).

The paper is organized as follows. In section 2 we describe the background model, introduce the energy-momentum tensor for the inhomogeneities, perturb the background
metrics by these inhomogeneities and obtain the master equation for the gravitational potential of this system. The gravitational potentials as well as equations of
motion for test masses forming the Hubble flows are found in sections 3 and 4 for conformally flat and hyperbolic spaces, respectively. The main results are summarized
in concluding section 5.


\section{\label{sec:2}Scalar perturbations of FRW Universe}

\setcounter{equation}{0}

To start with, we consider a homogeneous isotropic Universe described by the Friedmann-Robertson-Walker (FRW) metrics
\be{2.1}
ds^2=a^2\left(d\eta^2-\gamma_{\alpha\beta}dx^{\alpha}dx^{\beta}\right)=a^2\left(d\eta^2-\frac{\delta_{\alpha\beta}dx^\alpha
dx^\beta}{\left[1+\cfrac{1}{4}\mathcal K\left(x^2+y^2+z^2\right)\right]^2}\right)\, ,
\ee
where $\mathcal K=-1,0,+1$ for open, flat and closed Universes, respectively. The Friedmann equations for this metrics in the case of the $\Lambda$CDM model read
\be{2.2}
\frac{3\left({\mathcal H}^2+\mathcal K\right)}{a^2}=\kappa\overline{T}_{0}^0+\Lambda
\ee
and
\be{2.3}
\frac{2{\mathcal H}'+{\mathcal H}^2+\mathcal K}{a^2}=\Lambda\, ,
\ee
where ${\mathcal H}\equiv a'/a\equiv (da/d\eta)/a$ and $\kappa\equiv 8\pi G_N/c^4$ ($c$ is the speed of light and $G_N$ is the Newton's gravitational constant).
Hereafter, the Latin indices $i,k,=0,1,2,3$ and the Greek indices $\alpha,\beta=1,2,3$. $\overline T^{ik}$ is the energy-momentum tensor of the averaged pressureless
dustlike matter\footnote{For the late stages of the Universe evolution, we neglect the contribution of radiation.}. For such matter, the energy density $\overline
T^{0}_{0} =\overline \rho c^2/a^3$ is the only non-zero component. $\overline \rho$ is a constant which we define below. It is worth noting that in the case $\mathcal K
=0$ the comoving coordinates $x^{\alpha}$ may have a dimension of length, but then the conformal factor $a$ is dimensionless, and vice versa. However, in the cases
$\mathcal K=\pm 1$ the dimension of $a$ is fixed. Here, $a$ has a dimension of length and $x^{\alpha}$ are dimensionless. For consistency, we shall follow this
definition for $\mathcal K=0$ as well. To make reading easier, we insert the following table where explain the dimensions adopted in our paper:

\vspace{1cm}

\
\begin{tabular}{|c|c|c|}

\hline

\ & \ & \

\\

{\bf\large quantity} & {\bf\large symbol} & {\bf\large dimension} \\

\ & \ & \

\\

\hline

\ & \ & \

\\

scale factor & $a$ & L (length) \\

\ & \ & \

\\

rest mass density in comoving frame & $\rho$ & M (mass) \\

\ & \ & \

\\

average rest mass density in comoving frame & $\overline{\rho}$ & M (mass) \\

\ & \ & \

\\

\hline
\end{tabular}

\

\vspace{0.5cm}

Conformal time $\eta$ and synchronous time $t$ are connected as $cdt=a d\eta$. Therefore, eqs. \rf{2.2} and \rf{2.3}, respectively, take the form
\be{2.4}
H^2=\left(\frac{\dot a}{a}\right)^2=\frac{\kappa\overline\rho c^4}{3a^3}+\frac{\Lambda c^2}{3}-\frac{\mathcal K
c^2}{a^2}=H_0^2\left(\Omega_{M}\frac{a_0^3}{a^3}+\Omega_{\Lambda}+\Omega_{\mathcal K}\frac{a_0^2}{a^2}\right)\, ,
\ee
and
\be{2.5}
\frac{\ddot a}{a}=-\frac{\kappa\overline\rho c^4}{6a^3}+\frac{\Lambda c^2}{3}=H_0^2\left(-\frac{1}{2}\Omega_{M}\frac{a_0^3}{a^3}+\Omega_{\Lambda}\right)\, ,
\ee
where $a_0$ and $H_0$ are the values of the conformal factor $a$ and the Hubble "constant" $H\equiv \dot a/a\equiv (da/dt)/a$ at the present time $t=t_0$ (without loss
of generality, we can put $t_0=0$), and we introduced the standard density parameters:
\be{2.6} \Omega_M=\frac{\kappa\overline\rho c^4}{3H_0^2a_0^3},\quad \Omega_{\Lambda}=\frac{\Lambda c^2}{3H_0^2},\quad\Omega_{\mathcal K}=-\frac{\mathcal K
c^2}{a_0^2H_0^2}\, ,\ee
therefore
\be{2.7}
\Omega_M+\Omega_{\Lambda}+\Omega_{\mathcal K}=1\, .
\ee

The spatial part of the metrics \rf{2.1} can be also written in spherical coordinates:
\be{2.8}
dl^2=\gamma_{\alpha\beta}dx^{\alpha}dx^{\beta}=d\chi^2+\Sigma^2(\chi ) d\Omega^2_2\, ,
\ee
where
\be{2.9}
\Sigma (\chi)=\left\{
\begin{array}{ccc}
\sin \chi\, ,  & \chi \in [0,\pi] & \mbox{for} \quad {\mathcal K}=+1 \\
\chi\, ,  & \chi \in [0,+\infty) & \mbox{for} \quad {\mathcal K}=0 \\
\sinh \chi\, ,  & \chi \in [0,+\infty) & \mbox{for} \quad {\mathcal K}=-1
\end{array} \right .
\ee

In our paper, we consider the stage of the Universe evolution, which is much latter than the recombination time. At this stage, the formation of inhomogeneities (stars,
galaxies, clusters of galaxies) has been generally completed. As we mentioned in Introduction, the cell of statistical homogeneity/uniformity size is of the order of 150
Mpc \cite{Labini}-\cite{Wilt2}. On much bigger scales
the Universe is well described by the $\Lambda$CDM model with matter
mainly in the form of dark matter plus the cosmological constant. Here, dark matter is well simulated by a pressureless perfect fluid, i.e. dust, and the hydrodynamical
approach provides the adequate description of the model. On scales smaller than the cell of uniformity size, e.g., less than 150 Mpc, and on late stages of evolution,
the Universe is highly inhomogeneous and structured. We can see isolated galaxies, which form clusters and superclusters. The observations also strongly argue in favor
of dark matter concentrated around these structures. Obviously, these visible and invisible isolated inhomogeneities can not be represented in the form of liquid.
Therefore, hydrodynamics is not appropriate to describe their behavior on the considered scales. We need to create a mechanical approach where dynamical behavior is
defined by gravitational potentials. To perform it, we start with the energy-momentum tensor of non-interacting (except for the gravity) randomly distributed particles
(inhomogeneities, in our case) \cite{Landau}:
\ba{2.10}
T^{ik} &=&\sum_p\frac{m_p c^2}{(-g)^{1/2}[t]}\frac{dx^i}{ds}\frac{dx^k}{ds}\frac{ds}{cdt}\delta({\bf r}-{\bf r}_p)=
\sum\limits_p\frac{m_p c^2}{(-g)^{1/2}[\eta]}\frac{dx^i}{ds}\frac{dx^k}{ds}\frac{ds}{d\eta}\delta({\bf r}-{\bf r}_p)\nn\\
&=&\sum\limits_p\frac{m_p c^2}{(-g)^{1/2}[\eta]}\frac{dx^i}{d\eta}\frac{dx^k}{d\eta}\frac{d\eta}{ds}\delta({\bf r}-{\bf r}_p)\, ,
\ea
where $m_p$ is the mass of p-th inhomogeneity and $[t]$ and $[\eta]$ indicate that the determinant is calculated from the metric coefficients defined with respect to
synchronous $t$ or conformal $\eta$ times. In the rest of this section we deal with conformal time $\eta$ and we drop for simplicity the symbol $[\eta]$. It is worth
noting that for the Universe filled by inhomogeneities,
eq. \rf{2.10} describes the true
energy-momentum tensor of matter at any cosmic scales.

In the $\Lambda$CDM model, the main contributions come from the cosmological constant and the nonrelativistic matter. Therefore, the peculiar velocities should be much
less than the speed of light:
\be{2.11}
\frac{dx^{\alpha}}{d\eta}
=a\frac{dx^{\alpha}}{dt} \frac{1}{c}\ll 1\, .
\ee
Therefore, we can assume that $T^{00}$ is the only non-zero component of the energy-momentum tensor:
\be{2.12}
T^{00}=\sum_p \frac{m_p c^2}{(-g_{00}g)^{1/2}} \delta({\bf r}-{\bf r}_p) = \frac{\sqrt{\gamma}\rho c^2}{(-g_{00}g)^{1/2}}\, ,
\ee
where $\gamma$ is the determinant of the metrics $\gamma_{\alpha\beta}$ and we introduce the rest mass density
\be{2.13}
\rho=\frac{1}{\sqrt{\gamma}}\sum_p m_p \delta({\bf r}-{\bf r}_p)\, .
\ee
After averaging \rf{2.12} over all space, we get $\overline{T}^{00}=\overline{\rho}c^2/a^5$, where $\overline \rho$ is the average rest mass density \rf{2.13} and we use
the unperturbed metrics \rf{2.1}. Therefore, $\overline{T}_{0}^0=\overline{\rho}c^2/a^3$ as we have mentioned above.

Obviously, the inhomogeneities in the Universe result in scalar perturbations of the metrics \rf{2.1}. In the conformal Newtonian gauge, such perturbed metrics is
\cite{Mukhanov-book,Rubakov-book}
\be{2.14}
ds^2\approx a^2\left[(1+2\Phi)d\eta^2-(1-2\Psi)\gamma_{\alpha\beta}dx^{\alpha}dx^{\beta}\right]\, ,
\ee
where scalar perturbations $\Phi$ and $\Psi$ depend on all space-time coordinates $\eta,x,y,z$ and satisfy equations
\be{2.15}
\triangle\Psi-3{\mathcal H}(\Psi'+{\mathcal H}\Phi)+3\mathcal K \Psi=\frac{1}{2}\kappa a^2\delta T_{0}^0\, ,
\ee
\be{2.16}
\frac{\partial}{\partial x^{\beta}}(\Psi'+{\mathcal H}\Phi)=\frac{1}{2}\kappa a^2\delta T_{\beta}^0=0\, ,
\ee
\ba{2.17}
&{}&\left[\Psi''+{\mathcal H}(2\Psi+\Phi)'+\left(2{\mathcal H}'+{\mathcal H}^2\right)\Phi+\frac{1}{2}\triangle(\Phi-\Psi)-\mathcal
K\Psi\right]\delta^{\alpha}_{\beta}\nn\\
&-&\frac{1}{2}\gamma^{\alpha\sigma}\left(\Phi-\Psi\right)_{;\sigma;\beta}=-\frac{1}{2}\kappa a^2\delta T_{\beta}^{\alpha}=0\, ,
\ea
where the Laplace operator $\triangle$ and the covariant derivatives are defined with respect to the metrics $\gamma_{\alpha\beta}$. The condition $\delta T_{\beta}^0=0$
follows from the nonrelativistic nature of the considered matter, i.e. $|\delta T_{\beta}^0| \ll \delta T_{0}^0$, and we can drop $\delta T_{\beta}^0$ with respect to
$\delta T_{0}^0$. To clarify this point, we want to stress that according to Eqs. \rf{2.15} and \rf{2.16}, both $\delta T_{0}^0$ and $\delta T_{\beta}^0$ contribute to
the gravitational potential $\Phi$. However, due to the condition \rf{2.11}, peculiar velocities of inhomogeneities are nonrelativistic and the contribution of $\delta
T_{\beta}^0$ is negligible compared to that of $\delta T_{0}^0$. In other words, account of $\delta T_{\beta}^0$ is beyond the accuracy of the model. This approach is
fully consistent with \cite{Landau} where it is shown that the nonrelativistic gravitational potential is defined by the positions of the inhomogeneities but not by
their velocities (see Eq. (106.11) in this book). In the case of an arbitrary number of dimensions, a similar result was obtained in \cite{EZ3}. The perturbed matter
remains nonrelativistic (pressureless) that results in the condition $\delta T_{\beta}^{\alpha}=0$. Below, we define the accuracy, with which the latter statement holds.
Following the standard argumentation (see, e.g., \cite{Mukhanov-book,Rubakov-book}), we can put $\Phi=\Psi$, then the system of above equations reads
\be{2.18}
\triangle\Phi-3{\mathcal H}(\Phi'+{\mathcal H}\Phi)+3\mathcal K \Phi=\frac{1}{2}\kappa a^2\delta T_{0}^0\, ,
\ee
\be{2.19}
\frac{\partial}{\partial x^{\beta}}(\Phi'+{\mathcal H}\Phi)=0\, ,
\ee
\be{2.20}
\Phi''+3{\mathcal H}\Phi'+\left(2{\mathcal H}'+{\mathcal H}^2\right)\Phi-\mathcal K \Phi=0\, .
\ee
From eq. \rf{2.19} we get
\be{2.21}
\Phi(\eta,{\bf r})=\frac{\varphi({\bf r})}{c^2a(\eta)}\, ,
\ee
where $\varphi({\bf r})$ is a function of all spatial coordinates and we have introduced $c^2$ in the denominator for convenience (for such normalization, $\varphi$ has
the dimension $(c^2)\times (length)$). The function $\Phi$ is defined up to an arbitrary additive function $f(\eta)$ which does not depend on the spatial coordinates.
Therefore, this function $f$ is not related to the inhomogeneities and we can drop it. Below, we shall see that $\varphi({\bf r})\sim 1/r$ in the vicinity of an
inhomogeneity, and the nonrelativistic gravitational potential $\Phi(\eta,{\bf r})\sim 1/(a r)=1/R$, where $R=ar$ is the physical distance. Hence, $\Phi$ has the correct
Newtonian limit near the inhomogeneities. Substituting the expression \rf{2.21} into eq. \rf{2.18}, we get the following equation for $\varphi$:
\be{2.22}
\triangle\varphi+3\mathcal K\varphi=\frac{1}{2}\kappa c^2a^3\delta T_{0}^0\, .
\ee
The left hand side of this equation is independent of $\eta$, hence, the right hand side also should not depend on $\eta$. This is possible only if $\delta
T_{0}^0\sim1/a^3$. Let us now define the conditions of implementation of this statement. According to eq. \rf{2.12}, we have
\be{2.23}
T_{0}^0=\frac{\sqrt{\gamma}\rho c^2\sqrt{g_{00}}}{\sqrt{-g}}\, .
\ee
Taking into account the perturbed metrics \rf{2.14}, we get in the first approximation
\be{2.24}
\delta T_{0}^0=\frac{\delta\rho c^2}{a^3}+\frac{3\overline{\rho}c^2\Phi}{a^3}=\frac{\delta\rho c^2}{a^3}+\frac{3\overline{\rho}\varphi}{a^4}\, ,
\ee
where $\delta\rho$ is the difference between real and average rest mass densities:
\be{2.25}
\delta\rho = \rho-\overline\rho\, .
\ee
In the right hand side of eq. \rf{2.24}, the second term is proportional to $1/a^4$ and should be dropped because we consider the nonrelativistic matter. This is the
accuracy of our approach. Hence, the perturbation of the energy-density reads
\be{2.26}
\delta T_{0}^0=\frac{\delta\rho c^2}{a^3}\, .
\ee
Finally, eq. \rf{2.22} takes the form
\be{2.27}
\triangle\varphi+3\mathcal K\varphi=4\pi G_N (\rho-\overline\rho)\, .
\ee
This is our master equation where the appearance of $\overline \rho$ plays a crucial role. It is worth mentioning that $\varphi$ is not a "physical" gravitational
potential, and eq. \rf{2.27} looks only formally as the usual Poisson equation. Here, $\rho$ and $\overline\rho$ are comoving local and average rest mass densities,
respectively, which do not depend on time. The connection between $\varphi$ and the physical potential $\Phi$ is given by eq. \rf{2.21}.

We now turn to eq. \rf{2.20}. Taking into account the relations
\be{2.28}
\Phi'=-\frac{\varphi a'}{c^2a^2}=-{\mathcal H}\Phi,\quad\Phi''=-{\mathcal H}'\Phi+{\mathcal H}^2\Phi\, ,
\ee
we can rewrite this equation in the following form:
\be{2.29}
\left({\mathcal H}'-{\mathcal H}^2-\mathcal
K\right)\Phi=0\, .
\ee
On the other hand, eqs. \rf{2.2} and \rf{2.3} show that
\be{2.30}
\frac{2\left({\mathcal H}'-{\mathcal H}^2-\mathcal K\right)}{a^2}=-\kappa \overline{T}_{0}^0\sim\frac{1}{a^3}\, .
\ee
Therefore, for $\Phi$ from \rf{2.21}, eq. \rf{2.20} is satisfied with the adopted accuracy $O\left(1/a^4\right)$.

It is worth noting that for an arbitrary perfect fluid with the equation of state $p=\omega\varepsilon$, eq. \rf{2.30} reads
\be{2.31}
\frac{2\left({\mathcal H}'-{\mathcal H}^2-\mathcal K\right)}{a^2}=-\kappa (\varepsilon+p)=-\kappa (1+\omega)\varepsilon\, .
\ee
For the cosmological constant, the right hand side is equal to zero. However, in the case $\omega \ne -1$, the right hand side behaves as $1/a^{3(1+\omega)}$. The demand
$\varepsilon \sim O(1/a^n),\ n\ge 3$ leads to the condition $\omega \ge 0$. Therefore, homogeneous quintessence ($-1<\omega<0$) and phantom matter ($\omega<-1$) can not
be an alternative to the cosmological constant. In the papers \cite{ZWS,SWZ}, it was also pointed out that the quintessence has to be inhomogeneous.

Coming back to the perturbed metrics \rf{2.14}, we can write it now in the form
\be{2.32} ds^2\approx\left(1+2\Phi\right)c^2dt^2-a^2\left(1-2\Phi\right)
\gamma_{\alpha\beta}dx^{\alpha}dx^{\beta}\, .
\ee
We shall use this metrics for investigation of motion of nonrelativistic test masses.
The action for a test body of the mass $m$ can be written in the following form \cite{Landau}:
\be{2.34}
S=-mc\int ds\approx-mc\int\left\{\left(1+2\Phi\right)c^2-a^2\left(1-2\Phi\right)v^2\right\}^{1/2}dt\, ,\quad
v^2=\gamma_{\alpha\beta}\dot{x}^{\alpha}\dot{x}^{\beta}\, ,
\ee
where $v$ is a comoving peculiar velocity which has the dimension $(time)^{-1}$.
Hence, the corresponding Lagrange function has the form
\be{2.35}
L\approx-mc^2\left\{1+2\Phi -a^2\frac{v^2}{c^2}(1-2\Phi)\right\}^{1/2}\approx-mc^2\left(1+\frac{\varphi}{ac^2}
-\frac{a^2v^2}{2c^2}\right)=-mc^2-\frac{m\varphi}{a}+
\frac{ma^2v^2}{2}\, ,
\ee
where we dropped the term $O\left(1/c^2\right)$. For nonrelativistic test masses, we can also drop the term $mc^2$. To get the Lagrange equations, we need now the form
of the gravitational potential $\varphi$ which is the solution of eq. \rf{2.27}. Below, we investigate eq. \rf{2.27} separately for the cases $\mathcal K = 0$ and
$\mathcal K = \pm 1$.


\section{\label{sec:3}Hubble flows in conformally flat space ($\mathcal K=0$) }

\setcounter{equation}{0}

In this case eq. \rf{2.27} reads
\be{3.1}
\triangle\varphi=4\pi G_N (\rho-\overline\rho)\, ,
\ee
where the Laplace operator $\triangle = \sum_{\alpha=1}^3 \partial^2/\left(\partial x^{\alpha}\right)^2$ and the rest mass density $\rho$ is defined by \rf{2.13} where
$\gamma=1$. Obviously, the presence of $\overline \rho$ destroys the superposition principle (it is hardly possible to single out the contribution of each of the
inhomogeneities to the average density $\overline\rho$). There is also the well known problem related to the solution of the Poisson equation for the infinite space with
homogeneous energy density (the Neumann-Seeliger paradox) or infinite number of gravitating masses. To find a finite solution of such equation, we need to assume the
form of the spatial distribution of these masses. Therefore, to avoid this problem in our case, we suppose that in the vicinity of each inhomogeneity, the gravitational
potential is defined by the mass of this inhomogeneity and is not affected by other masses. We consider a simplified version where  the inhomogeneities are approximated
by point-like masses, which do not interact gravitationally with each other. We can easily generalize this picture to the case where some of inhomogeneities form a
gravitationally bound system. In this case, we consider such system as one point-like mass concentrated in the center of mass of this system.
Further, we assume that each point-like mass $m_{0{i}}$ (here, we introduce the subscript $0$ to differ these gravitating masses from a test mass $m$) is surrounded by
an empty sphere of the radius $r_{0{i}}$ defined below and this sphere, in turn, is embedded in a medium{\footnote{Obviously, in the region of space where the energy
density coincides with  the average energy density $\overline\rho c^2/a^3$,
we restore the unperturbed Friedmann Universe with the ideal cosmological medium characterized by $\overline\rho$. This ideal cosmological medium takes into account the
average effect of all inhomogeneities. It seems tempting to solve the problem  of missing dark matter in the local Universe  with the help of this cosmological medium
which may play the role of a "smooth ocean" of dark matter suggested in \cite{Kar2012}. As we shall see below, test masses outside of the spheres (i.e. for $r\ge r_0$)
form Hubble flows which are not affected by the gravitational attraction of the inhomogeneities. The gravitational field of an inhomogeneity stops to act outside of this
sphere. For this reason, we refer to these surfaces as the spheres of local gravity, and the radius $r_0$ is the radius of local gravity.} with the rest mass density
$\overline\rho$ (see Figure \ref{K=0}). Below, we shall demonstrate that such supposition of the spatial distribution of matter provides the finiteness of the
gravitational potential at any point of space and for an arbitrary number of inhomogeneities.

\begin{figure}[hbt]
\center{\includegraphics[width=9cm,height=7cm]{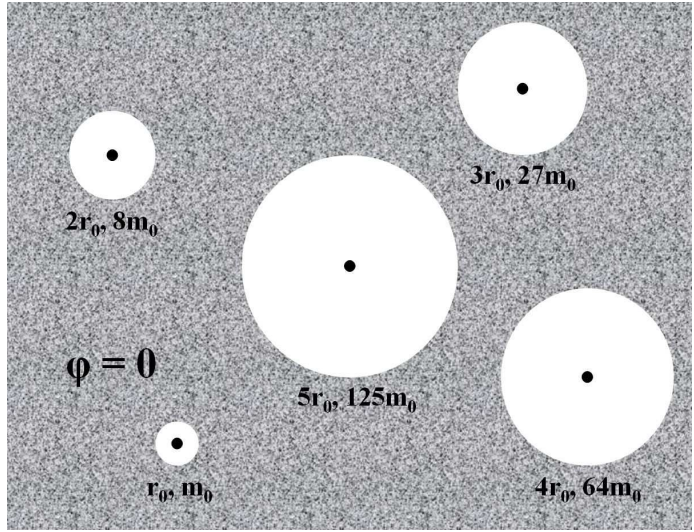}}
\caption{ For the model $\mathcal{K}=0$, we depict schematically an example of the large-scale structure with five gravitationally unbound inhomogeneities (e.g., groups
of galaxies). Each of inhomogeneities is surrounded by the vacuum sphere of local gravity with the Poisson equation \rf{3.2} inside of this sphere. The radius $r_{0i}$
of the sphere and the mass $m_{0i}$ of the inhomogeneity are related by eq. \rf{3.6}. All these spheres are embedded into ideal cosmological medium (the shaded region),
where the gravitational potential is absent. Inside of this region, test masses follow the Hubble flows diluted by peculiar velocities (see eq. \rf{3.13}). For
definiteness, we consider the spheres of local gravity of radii $r_{01}=r_0, r_{02}=2r_0, r_{03}=3r_0, r_{04}=4r_0$ and $r_{05}=5r_0$.\label{K=0}}
\end{figure}

Therefore, inside of each of these spheres, the Poisson equation \rf{3.1} reads
\be{3.2} \triangle\varphi=\frac{1}{r}\frac{d^2}{dr^2}(r\varphi)=4\pi G_N[m_0\delta({\bf r})-\overline{\rho}]\, , \quad r\le r_{0}\, , \ee
where, for simplicity, we omit the subscript $i$ for the mass $m_0$, the radius $r_0$ and the gravitational potential $\varphi$. In the cosmological medium, the Poisson
equation is
\be{3.3} \triangle\varphi =0\, , \quad r\ge r_0\, . \ee
For this equation we choose the trivial solution
\be{3.4}
\varphi\equiv 0\quad \Rightarrow\quad \frac{d\varphi}{dr}=0\, , \quad r\ge r_0\, .
\ee
It can be easily seen that the solution of eq. \rf{3.2} with the Newtonian limit at $r=0$, which is smoothly matched with \rf{3.4} at $r=r_0$, is
\be{3.5}
\varphi=-\frac{G_Nm_0}{r}-\frac{G_Nm_0}{2r_0^3}r^2+\frac{3G_Nm_0}{2r_0}\, ,\quad r\le r_0\, ,
\ee
where the matching condition gives
\be{3.6}
\quad r_0=\left(\frac{3m_0}{4\pi\overline{\rho}}\right)^{1/3}\, .
\ee
Eq. \rf{3.6} shows that $r_0$ is just the radius of a sphere of the rest mass density $\overline\rho$ and the total mass $m_0 = 4\pi r_0^3\overline{\rho}/3$. The second
term in the right hand side of eq. \rf{3.5} is divergent at large distances:
\be{3.7}
\frac{2\pi G_N\overline\rho}{3}r^2=\frac{G_Nm_0}{2r_0^3}r^2 \to +\infty\quad \mbox{for}\quad r\to +\infty\, .
\ee
This was the reason of the cutoff of the potential \rf{3.5} at $r=r_0$. The nonrelativistic gravitational potential in the Schwarzschild-de Sitter model is not free from
the similar quadratic divergence. It is worth mentioning that $r_0$ and $\overline\rho$ are comoving quantities which do not depend on time. $r_0$ is dimensionless,
$\overline\rho$ has the dimension of mass and $\varphi$ has the dimension $\left(c^2\right)\times (length)$. The physical rest mass density is
$\overline{\rho}_{ph}=\overline{\rho}/a^3$. It makes sense to estimate at present time ($a=a_0$) the physical radius of the sphere of local gravity $R_0(a=a_0)\equiv
\overline{R}_0=a_0r_0=a_0\left[3m_0/(4\pi\overline{\rho})\right]^{1/3}=\left[2G_Nm_0/\left(H_0^2\Omega_M\right)\right]^{1/3}$ for our group of galaxies with $m_0\approx
1.9\times 10^{12} M_{\bigodot}\approx 3.8\times 10^{45}$g \cite{Kar2012}. According to the seven-year WMAP observations \cite{7WMAP}, $H_0\approx70\,
\mbox{km/sec/Mpc}\approx 2.3\times 10^{-18}\mbox{sec}^{-1}$, $\Omega_M\approx0.27$ and $\Omega_{\Lambda}\approx0.73$. Hence, $\overline{R}_0 \approx 7.1\times
10^{24}\mbox{cm}\approx 2.3$ Mpc.


Now let us investigate motion of a test mass $m$ in the considered background. The equation of motion is defined by the Lagrange function \rf{2.35} where the
gravitational potential is given by eqs. \rf{3.4} and \rf{3.5}. The Lagrange equation is
\be{3.8}
\frac{d}{dt}\left(a^2{\bf v}\right)=-\frac{1}{a}\frac{\partial \varphi}{\partial {\bf r}}\, ,
\ee
where for the flat comoving space ${\bf v}=d{\bf r}/dt$. According to eq. \rf{3.4}, in the cosmological medium, the gravitational potential $\varphi\equiv 0$ and
integration of \rf{3.8} gives
\be{3.9} {\bf v}=\frac{{\bf c}_1}{a^2},\quad {\bf r}={\bf c}_1\int\limits_{t_0}^t\frac{1}{a^2}dt+{\bf c}_2\, , \ee
where ${\bf c}_1$ and ${\bf c}_2$ are constants of integration defined with respect to the initial time $t_0$. ${\bf v}$ is the peculiar velocity in the comoving
coordinates. The physical distance is ${\bf R} =a{\bf r}$. Therefore, the physical velocity ${\bf V}\equiv d{\bf R}/dt=d(a{\bf r})/dt$ is related to ${\bf v}$ as
follows:
\be{3.10}
{\bf V}=\frac{\dot{a}}{a}{\bf R}+a{\bf v}\, ,
\ee
This equation shows that the physical peculiar velocity is ${\bf v}_{ph}=a{\bf v}$. With the help of the relation \rf{3.10}, we can rewrite the Lagrange function
\rf{2.35} with respect to the physical quantities ${\bf V}$ and ${\bf R}$:
\be{3.11}
L=-mc^2-\frac{m\varphi}{a}+\frac{mV^2}{2}+
\frac{m\dot{a}^2R^2}{2a^2}-\frac{m\dot{a}}{a}({\bf VR}) \, .
\ee
The corresponding Lagrange equation is
\be{3.12}
\frac{d}{dt}\left({\bf V}-\frac{\dot{a}}{a}{\bf R}\right)=-\frac{1}{a}\frac{\partial
\varphi}{\partial {\bf R}}+\frac{\dot{a}^2}{a^2}{\bf R}-\frac{\dot{a}}{a}{\bf V}\, .
\ee
In the case $\varphi=0$, i.e. in the cosmological medium ($r\ge r_0$), we can easily integrate this equation:
\be{3.13} {\bf V}=\frac{\dot{a}}{a}{\bf R}+\frac{{\bf c}_1}{a},\quad {\bf R}=a{\bf c}_1\int\limits_{t_0}^t\frac{1}{a^2}dt+a{\bf c}_2\, . \ee
Therefore, in the absence of the gravitational potential, the physical velocity consists of two parts. They are the Hubble velocity which is "diluted" by the peculiar
velocity. Obviously, for $\varphi=0$, the Lagrange equation defines the Hubble flows. Here, eq. \rf{3.12} in the case of radial motion can be rewritten in the
form\footnote{For radial motion, the equation $a\ddot{R}=\ddot{a}R \; \Rightarrow \; d\left(a^2\dot{r}\right)/dt=0$ is equivalent to eq. \rf{3.8} if $\varphi=0$.}
\be{3.14}
{\ddot{R}}=\frac{\ddot{a}}{a} R=\left(-\frac{4\pi G_N\overline\rho}{3a^3}+\frac{\Lambda c^2}{3}\right)R
=H_0^2\left(-\frac{1}{2}\Omega_{M}\frac{a_0^3}{a^3}+\Omega_{\Lambda}\right)R\, ,
\ee
where
we use the Friedmann equation \rf{2.5}. It can be easily seen that in the cosmological medium ($r\ge r_0$) the Hubble flows exist even in the absence of the cosmological
constant. It is therefore not correct to say that the Hubble flows at small distances (a few megaparsecs) are due solely to the cosmological constant (see, e.g.,
\cite{CherninUFN2}). The reason for the Hubble flow is the global cosmological expansion of the Universe. However, the acceleration ($dV/dt >0$) is possible only in the
presence of the positive cosmological constant. The cosmological constant also reduces smearing of the Hubble flow by peculiar velocities \cite{Chernin2004}.

Obviously, in spherical coordinates $r,\theta=\pi/2,\psi$ the Lagrange function \rf{2.35} reads
\be{3.15}
L=-mc^2 -\frac{m\varphi}{a}+\frac{ma^2}{2}\left(\dot{r}^2+r^2\dot{\psi}^2\right)\, .
\ee
The corresponding Lagrange equations are
%
\be{3.16} \frac{d}{dt}\left(ma^2r^2\dot{\psi}\right)=0\quad \Rightarrow \quad \dot{\psi}=\frac{M}{ma^2r^2}\, \ee
and
\be{3.17}
\frac{d}{dt}\left(a^2\dot{r}\right)=-\frac{1}{a}\frac{\partial\varphi}{\partial r}+\frac{M^2}{m^2a^2r^3}\, ,
\ee
where $M=\mbox{const}$ is the angular momentum. Let us investigate the latter equation in the vicinity of the gravitating mass $m_0$, i.e. in the region $r\le r_0$ where
the gravitational potential is given by \rf{3.5}:
\be{3.18} \frac{d}{dt}\left(a^2\dot{r}\right)=-\frac{1}{a}\left(\frac{G_Nm_0}{r^2}-\frac{G_Nm_0}{r_0^3}r\right)+\frac{M^2}{m^2a^2r^3}\, .
\ee
After some algebra where we use the Friedmann equation \rf{2.5}, this equation takes the form
\be{3.19}
\ddot{R}=-\frac{G_Nm_0}{R^2}+\frac{M^2}{m^2R^3}+\frac{\Lambda c^2}{3}R\, , \quad R\le R_0=a r_0\, .
\ee
It is interesting to note that this equation contains the cosmological constant $\Lambda$, rather than the average density $\overline\rho$ in contrast to the expression
for the gravitational potential \rf{3.5}. Eq. \rf{3.19} can be also obtained from the Lagrange function \rf{3.11} with $\varphi$ from \rf{3.5}. Taking into account the
Friedmann equations \rf{2.4} and \rf{2.5}, this function reads
\be{3.20}
L=\frac{G_Nm_0m}{R}+\frac{m}{2}\frac{\Lambda c^2}{3}R^2+\frac{mV^2}{2}\, ,\quad V^2=\dot{R}^2+R^2\dot{\psi}^2\, ,\quad R\le R_0\, .
\ee
In classical mechanics, this Lagrange function describes motion of a test mass $m$ in the gravitational field with the potential
\be{3.21}
\tilde \varphi=-\frac{G_Nm_0}{R}-\frac{1}{2}\frac{\Lambda c^2}{3}R^2,\quad -\frac{\partial\tilde\varphi}{\partial R}=-\frac{G_Nm_0}{R^2}+\frac{\Lambda
c^2}{3}R\, , \quad R\le R_0\, .
\ee
Exactly this potential enters in the Schwarzschild-de Sitter metrics:
\be{3.22}
ds^2=\left(1+\frac{2\tilde\varphi}{c^2}\right)c^2dt^2-\left(1+\frac{2\tilde\varphi}{c^2}\right)^{-1}dR^2
-R^2\left(d\theta^2+\sin^2\theta d\psi^2\right)\, .
\ee
As it follows from \rf{3.21}, the derivative $-\partial\tilde\varphi/\partial R$ equals 0 at $R=R_*=\left[3G_Nm_0/\left(\Lambda c^2\right)\right]^{1/3}$. Therefore,
$R_0/R_*=\left(2\Omega_{\Lambda}/\Omega_M\right)^{1/3}(a/a_0)$. Since at present time ($a=a_0$) the parameters $\Omega_{\Lambda}$ and $\Omega_M$ satisfy the inequality
$\Omega_{\Lambda}\approx 0.73 > \Omega_M\approx 0.27$, we get that $R_*<\overline{R}_0$ (where $\overline{R}_0=R_0 (a=a_0)$). Hence, inside of the sphere of local
gravity, the attraction dominates for $R<R_*$ and repulsion dominates for $R_*<R<\overline{R}_0$. If $M=0$, then $\left.\ddot{R}\right|_{R_{*}}=0$ (see eq. \rf{3.19}),
and for the model $\mathcal{K}=0$, $R_*$ can be associated with the radius of the zero-acceleration surface. In some articles this surface is called the zero-velocity
surface \cite{Kar2012,Kar2008,Lynden-Bell,Kar et al} or the surface of zero gravity \cite{Chernin1}.
According to the observations \cite{Kar2012,Kar2008,Kar et al}, the radius of this surface is 0.96 Mpc for our group of galaxies. The calculated value $R_*=\overline
R_0\left(2\Omega_{\Lambda}/\Omega_M\right)^{-1/3}\approx 1.3$ Mpc is very close to the observed one.

Taking into account that $G_Nm_0/R_0^3=4\pi G_N\overline\rho/\left(3a^3\right)$, it can be easily seen that at $R=R_0$ eq. \rf{3.19} is 
matched to eq. \rf{3.14} (for radial motion $M=0$). Outside of the sphere of local gravity ($R\ge R_0$), test masses form the Hubble flows \rf{3.13}. Because
$\Omega_{\Lambda}>\Omega_M$, these flows are accelerated at the late stage of the Universe evolution (see eq. \rf{3.14}).


\section{\label{sec:4}Hubble flows in conformally hyperbolic space ($\mathcal K=-1$) }

\setcounter{equation}{0}

At the beginning of this section, we consider the general case of nonzero curvature ($\mathcal{K}=\pm 1$) and then dwell on the physically more interesting case of the
hyperbolic space ($\mathcal{K}=- 1$). To start with, we turn back to our master equation \rf{2.27}. For $\mathcal{K}\ne 0$, we can define a new function $\phi$:
\be{4.1}
\phi = \varphi+\frac{4\pi G_N\overline\rho}{3\mathcal K}\, ,
\ee
which satisfies the Helmholtz equation
\be{4.2} \triangle\phi+3\mathcal K\phi=4\pi G_N\rho\, , \ee
where $\rho$ is defined by eq. \rf{2.13}. It can be easily seen that the principle of superposition holds for eq. \rf{4.2}. Therefore, we can solve eq. \rf{4.2} for one
gravitating mass. Then, the solution for an arbitrary number of masses is the sum of the solutions for each separate mass (see below). So, let us consider one
gravitating mass which we denote as $m_0$. Without loss of generality, we may put this mass in the origin of coordinates. Because of spherical symmetry of the problem
and for the metrics \rf{2.8}, eq. \rf{4.2} in vacuum reads
\be{4.3}
\frac{1}{{\Sigma^2(\chi)}}\frac{d}{d\chi}\left({\Sigma^2(\chi)}\frac{d\phi}{d\chi}\right)+3\mathcal K\phi=0\, ,
\ee
where $\Sigma(\chi)$ is defined in \rf{2.9}.

\vspace{0.5cm}

First, we consider the case of the spherical space $\mathcal{K}=+1$ when $\Sigma(\chi)=\sin\chi$. The solution of this equation with the Newtonian limit at $\chi=0$ is
\be{4.4}
\phi=2C_1\cos\chi-G_Nm_0\left(\frac{1}{\sin\chi}-2\sin\chi\right)\, .
\ee
For any value of the constant of integration $C_1$, this solution (as well as its derivative $-d\phi/d\chi$) is divergent at $\chi=\pi$. This reflects the fact that the
surface area of the sphere which surrounds the gravitating source shrinks to zero for $\chi\to\pi$. The procedure of the cutoff which we explore in the flat space does
not work here because equations $\phi(\chi_0)=0$ and $d\phi/d\chi (\chi=\chi_0)=0$ are incompatible for $\forall\, \chi_0\in [0,\pi]$. Therefore, we disregard this
model.

\vspace{0.5cm}

Now, we consider the hyperbolic model with $\mathcal{K}=-1$ where $\Sigma(\chi)=\sinh\chi$. It can be easily seen that in this case the solution of eq. \rf{4.3}, which
has the Newtonian limit at $\chi\to 0$: $\phi \to -G_Nm_0/\chi$ and is finite at infinity: $\phi\to 0$ for $\chi \to +\infty$, reads
\be{4.5}
\phi=-G_Nm_0\frac{\exp(-2\chi)}{\sinh\chi}\, .
\ee
This formula demonstrates a number of advantages with respect to the flat space case. First, the presence of the exponential function enables us to avoid the
gravitational paradox. In some models, such exponential function was introduced by hand. In our model, it appears quite naturally. So, we do not need to introduce the
sphere of cutoff of the gravitational potential. The gravitating masses in this case can be distributed completely arbitrarily. Second, we can apply the principle of
superposition. For example, the function $\phi$ for all gravitating masses is
\be{4.6}
\phi=-G_N\sum\limits_i m_{0i}\frac{\exp(-2l_i)}{\sinh l_i}\, ,
\ee
where $l_i$ denotes the geodesic distance between the i-th mass $m_{0i}$ and the point of observation. Then, according to eq. \rf{4.1}, the gravitational potential of
this system is
\be{4.7} \varphi=-G_N\sum\limits_i m_{0i}\frac{\exp(-2l_i)}{\sinh l_i}+\frac{4\pi G_N\overline\rho}{3}\, .\ee

To investigate the Hubble flows in this model, we consider motion of a test mass $m$
in the vicinity of the gravitating mass $m_0$ where we can neglect the effect of other distant gravitating inhomogeneities, i.e. $i=1$ in the formula \rf{4.7}. In other
words, the function $\phi$ is given by eq. \rf{4.5}.
It is easy to verify that Lagrange equations in the equatorial plane $\theta=\pi/2$ for the considered case have the form
\be{4.8} \frac{d}{dt}\left(ma^2\sinh^2\chi\dot{\psi}\right)=0\quad \Rightarrow \quad \dot{\psi}=\frac{M}{ma^2\sinh^2\chi}\, \ee
and
\be{4.9}
\frac{d}{dt}\left(a^2\dot{\chi}\right)=-\frac{1}{a}\frac{\partial\phi}{\partial
\chi}+\frac{M^2}{m^2a^2}\frac{\cosh\chi}{\sinh^3\chi}\, .
\ee

It makes sense to estimate $\chi$ for the scales relevant to our model, i.e. for $R\lesssim 150$ Mpc. In the hyperbolic space, the physical distance is $R=a\chi$.
Obviously, $R\ll a\; \Rightarrow \; \chi\ll 1$. More precise limitation can be found if we know the value of the scale factor $a_0$ at present time which we can get from
the density parameter $\Omega_{\mathcal K=-1}=c^2/\left(a_0^2H_0^2\right)$. According to the seven-year WMAP observations, $\Omega_{\mathcal K=-1} < 8.4\times 10^{-3}$
(see section 4.3 in \cite{7WMAP}). Let us take $\Omega_{\mathcal K=-1}\sim 10^{-4}$. Then, for $H_0\approx 2.3\times 10^{-18} \mbox{sec}^{-1}$, we get $a_0\approx
4\times 10^5$Mpc. Therefore, $\chi\ll 1$ for all distances $R\ll 10^5$Mpc, and here we may approximate the function $\phi$ as follows: $\phi \approx -G_Nm_0/\chi$. For
such values of $\chi$, the Lagrange equation \rf{4.9} can be rewritten in the form
\be{4.10} \ddot R=\frac{\ddot a}{a}R-\frac{G_Nm_0}{R^2}+\frac{M^2}{m^2R^3} = H_0^2\left(-\frac{1}{2}\Omega_{M}\frac{a_0^3}{a^3}+\Omega_{\Lambda}\right)R
-\frac{G_Nm_0}{R^2}+\frac{M^2}{m^2R^3}\, , \ee
where we used the Friedmann equation \rf{2.5}. It is clear that the first term on the right hand side of this equation is responsible for the Hubble flow (see, e.g., eq.
\rf{3.14}). However, the second term (the gravitational attraction) works in the opposite direction trying to bind gravitationally the system. It makes sense to estimate
the distance where both of these terms are of the same order:
\be{4.11} \left|\frac{\ddot{a}}{a}\right|R\sim \frac{G_Nm_0}{R^2}\quad \Rightarrow\quad \left.R^3\right|_{a=a_0} \equiv {\overline{R}^{\, 3}_H}\sim
\left.\frac{G_Nm_0}{|{\ddot{a}/a}|}\right|_{a=a_0}=\frac{G_Nm_0}{H_0^2|\Omega_{\Lambda}-\Omega_M/2|}\, . \ee
Therefore, at present time, the Hubble flows begin to prevail over the gravitational attraction starting from $\overline{R}_H$. Obviously, $\overline{R}_H$ plays the
role of the radius of the zero-acceleration surface. It can be easily seen that the radius of the local gravity sphere in the flat space case
$\overline{R}_0=[2G_Nm_0/(H_0^2\Omega_M)]^{1/3}$ is related to $\overline{R}_H$ as follows: $\overline{R}_H = \overline{R}_0 \times
|(2\Omega_{\Lambda}/\Omega_M)-1|^{-1/3}$. For $\Omega_{\Lambda}\approx 0.73$ and $\Omega_M\approx 0.27$, we get for our group of galaxies $\overline{R}_H \approx 0.61\,
\overline{R}_0 \approx 1.4$ Mpc, which is rather close to the observed value 0.96 Mpc \cite{Kar2012}.


For $R>\overline{R}_H$, the Hubble flows are distorted by the gravitational attraction (it is the so called dragging effect \cite{Kar2012,Kar2008}). To quantify these
distortions, it is convenient to rewrite eq. \rf{4.10} in dimensionless units:
\be{4.12} \tilde a=\frac{a}{a_0},\quad \tilde t=H_0t,\quad \tilde R=\frac{R}{\overline{R}_H},\quad \tilde M^2=\frac{M^2}{H_0^2m^2\overline{R}_H^4}\, . \ee
Then, we get
\be{4.13} \frac{d^2\tilde R}{d\tilde t^2}=\left(-\frac{\Omega_{M}}{2\tilde a^3}+\Omega_{\Lambda}\right)\tilde R-\frac{\Omega_{\Lambda}-\Omega_{M}/2}{\tilde
R^2}+\frac{\tilde M^2}{\tilde R^3}\, . \ee
Without loss of generality, we may consider the case of the radial motion: $\tilde M=0$. Obviously, at the stages of the accelerating expansion of the Universe, it is
reasonable to study the Hubble flows at distances, where the right hand side of this equation is non-negative:
\be{4.14} \tilde R \ge \left(\frac{\Omega_{\Lambda}-\Omega_M/2}{\Omega_{\Lambda}-\Omega_M/\left(2\tilde a^3\right)}\right)^{1/3}\, , \quad \tilde a
>[\Omega_M/(2\Omega_{\Lambda})]^{1/3}\, , \ee
i.e. $\tilde R \ge 1\; \Rightarrow \; R \ge \overline{R}_H$ at the present time ($\tilde a =1$). Then, we can easily calculate the distance, at which the contribution of
the absolute magnitude of the gravitational attraction to the right hand side of eq. \rf{4.13} is $n\%$\footnote{At the late stage of the Universe evolution, the right
hand side of eq. \rf{4.13} (where for simplicity we put $\tilde M=0$) consists of positive and negative terms and is equal to zero at present time ($\tilde a=1$) for
$\tilde R=1$. In this case $n\%$ runs from 0 to infinity: $0\le n\%<+\infty$, which can be easily seen from the expression $n\%=\left(\tilde R^3-1\right)^{-1}100\%\, ,\;
\tilde R\ge 1$ obtained for $\tilde a=1$.}:
\be{4.15}
\tilde R_{n\%}=\left[\left(1+\frac{100\%}{n\%}\right)\frac{\Omega_{\Lambda}-\Omega_M/2}{\Omega_{\Lambda}-\Omega_M/(2\tilde a^3)}\right]^{1/3}\, .
\ee
For example, at present time, $\tilde R_{10\%}\approx 2.2$, $\tilde R_{5\%}\approx2.8$ and $\tilde R_{1\%}\approx4.7$.
For our group of galaxies it gives, respectively, $R_{10\%}\approx 3.1$ Mpc, $R_{5\%}\approx 3.9$ Mpc and $ R_{1\%}\approx 6.6$ Mpc. Therefore, in this model, the Hubble
flows are observed at sufficiently small distances much smaller than the cell of homogeneity size. To confirm this conclusion, we consider now some plots of the velocity
$V=dR/dt$ of a test body. In dimensionless units it reads
\be{4.16}
\tilde V =\frac{d\tilde R}{d\tilde t}=\frac{1}{\overline{R}_H H_0}V\, .
\ee
For the Hubble flow $V=H(t)R=H(t) \overline{R}_H \tilde R$ we get
\be{4.17}
\tilde V=\frac{H(t)}{H_0}\tilde R\, .
\ee
Hence, $\tilde V =\tilde R$ only at the present time $t=t_0$. Without loss of generality, we usually put $t_0=0\, \Rightarrow \, \tilde t_0=0$.

On Figure \ref{K=-1}, we depict the ratio $\tilde V/\tilde R$ for two cases (in both of them $\tilde M=0$). The upper (red) line is a result of numerical
integration\footnote{To perform such integration, we need to know the time dependence of the scale factor $a(t)$ which satisfies the Friedmann equation \rf{2.4}.
Obviously, because $\Omega_{\mathcal{K}}\ll 1$, we can drop the curvature term, and the solution of this equation is common for any value of $\mathcal{K}$. Then, in
dimensionless units, this equation reads $(\tilde a^{-1}d\tilde a/d\tilde t)^2=\Omega_M/\tilde a^3+\Omega_{\Lambda}$. The solution of this equation (with the boundary
condition $\tilde a=1$ at $\tilde t=\tilde t_0=0$) is
$$
\tilde
a=\left(\frac{\Omega_M}{\Omega_{\Lambda}}\right)^{1/3}\left[\left(1+\frac{\Omega_{\Lambda}}{\Omega_M}\right)^{1/2}\sinh\left(\frac{3}{2}\Omega_{\Lambda}^{1/2}\tilde
t\right)+\left(\frac{\Omega_{\Lambda}}{\Omega_M}\right)^{1/2}\cosh\left(\frac{3}{2}\Omega_{\Lambda}^{1/2}\tilde t\right)\right]^{2/3}\, .
$$
} of eq. \rf{4.13} for the pure Hubble flow case, i.e. when the second term (the gravitational attraction) in the right hand side of this equation is absent. The middle
(blue) line demonstrates how the gravitational attraction changes the Hubble flow, i.e. this is the ratio $\tilde V/\tilde R$ in the presence of the gravitational
attraction. The initial (at $\tilde t=\tilde t_0=0$) conditions are taken as follows: $\tilde R=1 \; \Rightarrow \; R=\overline{R}_H$ and $\tilde V/\tilde R =1\;
\Rightarrow \; V/R=H_0$. The dashed line is the eleven-percent deviation from the Hubble flow. Our calculations show that the blue line is entirely within this
deviation.

\begin{figure}[hbt]
\center{\includegraphics[width=9cm,height=6cm]{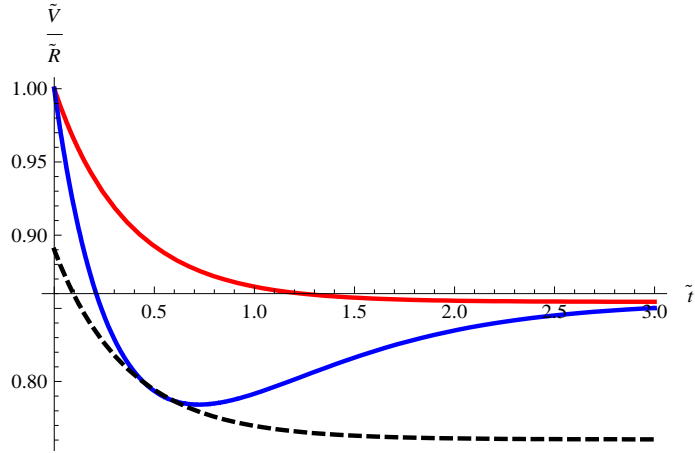}}
\caption{ Here, we depict the ratio  $\tilde V/\tilde R$ from the numerical solution of eq. \rf{4.13} without gravitational attraction (the upper/red line), i.e. in the
case of the pure Hubble flow, and  in the presence of this attraction (the middle/blue line). The value $\tilde t=0$ corresponds to the present time. The effect of
gravitational attraction does not exceed the eleven-percent deviation (the dashed line) from the Hubble flow. \label{K=-1}}
\end{figure}

On Figure \ref{M=0}, we depict the evolution of the velocity $\tilde V$ of a test mass with the distance $\tilde R$ in the case of the zero angular momentum $\tilde
M=0$. Solid and dashed lines correspond to presence and absence of the gravitational attraction, respectively. All lines start at $\tilde R=1\, \Rightarrow \,
R=\overline{R}_H$ with the following initial velocities: $\tilde V=1$ (blue lines), $\tilde V=2$ (green lines) and $\tilde V = 3$ (red lines). These graphs show that,
first, the effect of the gravitational attraction is not strong for $R>\overline{R}_H$ and, second, all lines go asymptotically to the dashed blue line which describes
the pure Hubble flow (i.e. the admixture of peculiar velocities is absent and the test body moves solely due to the expansion of the Universe). Peculiar velocities
result in deviation from the pure Hubble flow, and the plots demonstrate how fast test bodies with different initial values of $\tilde V$ join this flow. For our group
of galaxies, there is observational evidence \cite{Kar2008,Kar et al} that the peculiar velocities are of the order of 30 km/sec within 3 Mpc, i.e. the initial value
$\tilde V \sim 1.3$. It can be easily shown that in this case the corresponding solid line coincides with the pure Hubble flow already at $\tilde R = 2\; \Rightarrow \;
R \approx 2.8$ Mpc.

\begin{figure}[hbt]
\center{\includegraphics[width=9cm,height=7cm]{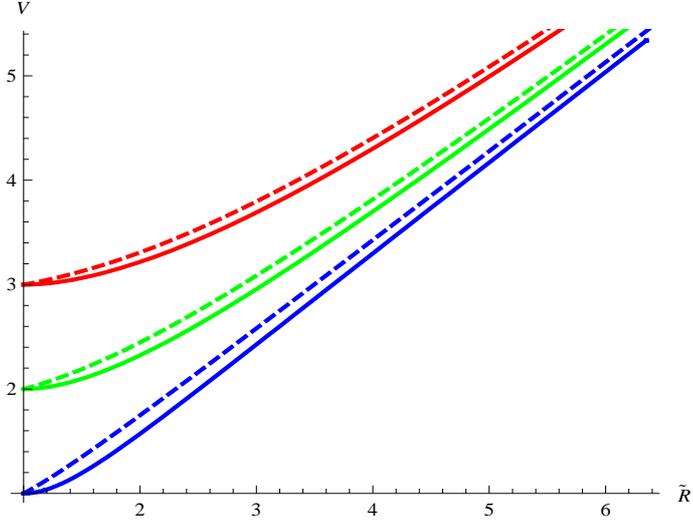}} \caption{This plot shows the evolution of the velocity $\tilde V$ of a test mass with the distance $\tilde R$.
All lines start at $\tilde R=1\, \Rightarrow \, R=\overline{R}_H$ with the following initial velocities: $\tilde V=1$ (blue lines), $\tilde V=2$ (green lines) and
$\tilde V = 3$ (red lines). Solid and dashed lines correspond to presence and absence of the gravitational attraction, respectively. The blue dashed line describes the
pure Hubble flow. For our group of galaxies, dimensional velocities and distances are $V\approx \tilde V \times 100$ km/sec and $R\approx \tilde R\times 1.4$
Mpc.}\label{M=0}
\end{figure}

On Figure \ref{M non 0}, we depict the effect of the angular momentum $\tilde M$ on the motion of  test masses. Solid and dashed lines correspond to presence and absence
of the gravitational attraction, respectively. All colored lines start at $\tilde R=1\, \Rightarrow \, R=\overline{R}_H$ with the zero initial velocity $\tilde V =0$ and
have the angular momentum $\tilde M=1$ (blue lines), $\tilde M=2$ (green lines) and $\tilde M=3$ (red lines). Black solid and dashed lines start with the initial
velocity $\tilde V=1$ at $\tilde R=1$ and have the zero angular momentum, i.e. the dashed black line here describes the pure Hubble flow. First, these colored graphs
demonstrate that test masses with the zero radial component of the velocity escape from the zero-acceleration surface outward due to the effect of the centrifugal force.
Second, the radial components of the velocity go asymptotically to the values for the pure Hubble flow. The smaller the angular momentum is, the sooner it happens. For
example, the blue lines with $\tilde M=1$ coincide with the corresponding black lines at $\tilde R \approx 2$.

\begin{figure}[hbt]
\center{\includegraphics[width=9cm,height=9cm]{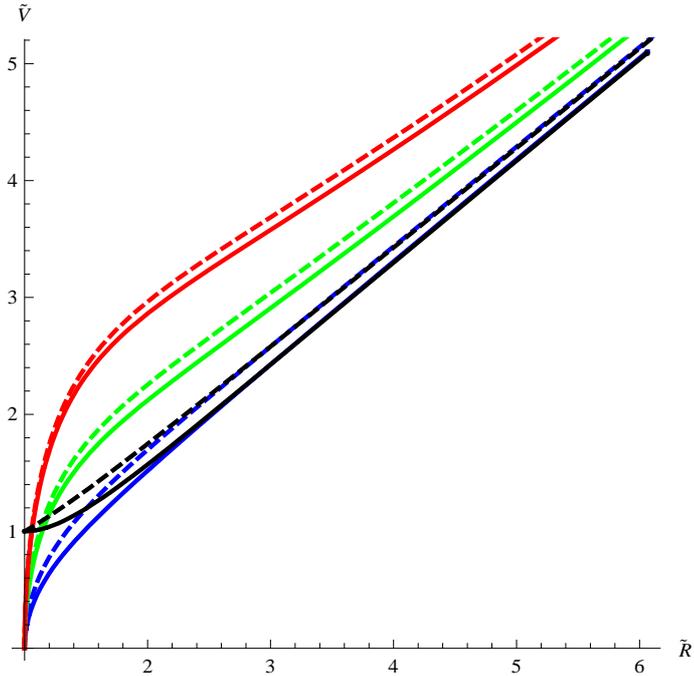}} \caption{This plot shows the effect of the angular momentum $\tilde M$ on motion of  test masses. Solid
and dashed lines correspond to presence and absence of the gravitational attraction, respectively. All colored lines start at $\tilde R=1\, \Rightarrow \,
R=\overline{R}_H$ with the zero initial velocity $\tilde V =0$ and have the angular momentum $\tilde M=1$ (blue lines), $\tilde M=2$ (green lines) and $\tilde M=3$ (red
lines). Here, $\tilde V$ is the radial component of the velocity of a test body. Black solid and dashed lines start with the initial velocity $\tilde V=1$ at $\tilde
R=1$ and have the zero angular momentum, i.e. the dashed black line describes the pure Hubble flow.}\label{M non 0}
\end{figure}

It is also of interest to draw the characteristic distances $\overline{R}_0=\left[2G_Nm_0/\left(H_0^2\Omega_M\right)\right]^{1/3}$, $R_*=\overline
R_0\left(2\Omega_{\Lambda}/\Omega_M\right)^{-1/3}$ and $\overline{R}_H = \overline{R}_0 \times |(2\Omega_{\Lambda}/\Omega_M)-1|^{-1/3}$ as the functions of the
gravitating mass $m_0$. Therefore, we depict these graphs on Figure \ref{distances}. Here, solid black (upper), blue (middle) and red (lower) lines correspond to
$\overline{R}_0$, $\overline{R}_H$ and $R_*$, respectively. These distances are given in Mpc. Vertical dashed orange (left) and green (right) lines correspond to our
Local Group of galaxies ($m_0\approx 1.9\times 10^{12} M_{\bigodot}$ \cite{Kar2012}) and the Virgo cluster ($m_0\approx 1.2\times 10^{15} M_{\bigodot}$ \cite{Virgo}),
respectively. This plot shows that at the present moment the radii of the zero acceleration sphere $\overline{R}_H$ for $\mathcal{K}=-1$ model and $R_*$ for
$\mathcal{K}=0$ model are very close to each other.

\begin{figure}[hbt]
\center{\includegraphics[width=9cm,height=9cm]{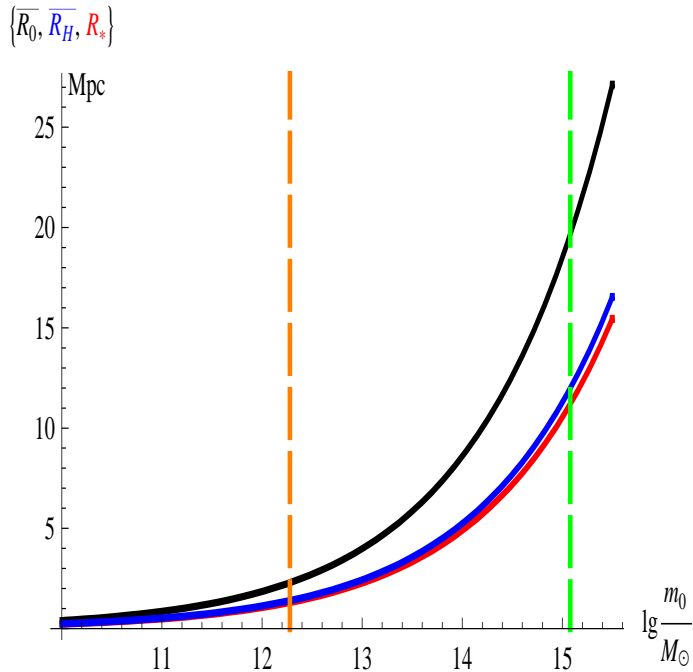}} \caption{This plot shows the characteristic distances $\overline{R}_0$ (black upper line), $\overline{R}_H$
(blue middle line) and $R_*$ (red lower line) as the functions of the gravitating mass $m_0$. Vertical dashed orange (left) and green (right) lines correspond to our
Local Group of galaxies and the Virgo cluster, respectively. }\label{distances}
\end{figure}

To conclude this section, we should note the following. It is obvious that the contribution of the constant term in the gravitational potential \rf{4.7} does not affect
the motion of nonrelativistic objects (see, e.g., eq. \rf{4.9}). However, this term contributes also in the metrics \rf{2.32} and at present time is of the order of
$8\pi G_N \overline\rho/\left(3a_0c^2\right)=\Omega_MH_0^2a_0^2/c^2 \sim 2\times 10^3$ where $a_0\sim 4\times 10^{5}$ Mpc. This large value does not contradict our
approach because we need to take into account the gravitational contribution \rf{4.6} of all inhomogeneities. To see it, let us average the i-th component in the first
term of \rf{4.7} over some finite volume $\mathcal{V}$:
\be{4.18} \overline\phi_i=\frac{4\pi}{\mathcal{V}} \int\limits_0^{\chi_0}\left[-G_Nm_{0i}\frac{\exp(-2\chi)}{\sinh\chi}\right]\sinh^2\chi d\chi =-\frac{4\pi
G_Nm_{0i}}{3\mathcal{V}}\left[1-\frac32\left(e^{-\chi_0}-\frac13e^{-3\chi_0}\right)\right]\, . \ee
Then, letting the volume go to infinity ($\chi_0 \to +\infty \;\Rightarrow\; \mathcal{V}\to+\infty$) and taking into account all gravitating masses, we obtain
\be{4.19} \overline\phi_{\mathrm{total}}=-\frac{4\pi G_N\overline\rho}{3}\, , \ee
where $\overline\rho = \lim\limits_{\mathcal{V}\to +\infty}\sum\limits_i m_{0i}/\mathcal{V}$.  Therefore, the averaged gravitational potential \rf{4.7} is equal to zero:
$\overline\varphi=0$. By the way, this is one more advantage of the considered model with $\mathcal K=-1$ as compared to the model with $\mathcal K=0$ where it can be
easily shown that the gravitational potential averaged over all spheres of local gravity has a non-vanishing negative value.

\section{Conclusion}

In this paper, we have considered our Universe at scales much less than the cell of homogeneity size which is approximately 150 Mpc. At such distances, our Universe is
highly inhomogeneous and averaged Friedmann approach does not work here. We need to take into account the inhomogeneities in the form of galaxies, groups of galaxies and
clusters of galaxies. All of them perturb the FRW metrics. We have investigated these perturbations in the weak-field limit where the $1/c^2$ correction term in the
metric coefficient $g_{00}$ defines the gravitational potential. The main goal was to get the expressions for this potential and to investigate the dynamical behavior of
test bodies (i.e. the dwarf galaxies) in the field of this potential. First, we have found the master equation \rf{2.27} for the
potential $\varphi({\bf r})$ which is conformally related to the physical gravitational potential $\Phi(\eta,{\bf r})$. Formally, this equation has the form of the
Poisson equation. A distinctive feature of this equation is the presence (in the right hand side) of the average comoving rest mass density of matter in the Universe.

Then, to solve this equation, we have studied in detail two types of models, namely with conformally flat $\mathcal{K}=0$ and conformally hyperbolic $\mathcal{K}=-1$
spaces. These models have both similarities and significant distinguishing features. For both of our models, gravitational potentials are finite at any point, including
spatial infinity\footnote{In the case of the spherical space $\mathcal{K}=+1$, the gravitational potential is divergent at $\chi=\pi$.}, and these solutions are valid
for an arbitrary number of gravitating masses. Note that the Schwarzschild-de Sitter solution was found for one gravitating source and the corresponding gravitational
potential is divergent at infinity. For both of these models, we have shown that at present time there is a distance from inhomogeneities at which the cosmological
expansion prevails over the gravitational attraction. For our group of galaxies
the radii of the zero-acceleration sphere ($R_*$ for $\mathcal{K}=0$ model and $\overline{R}_H$ for $\mathcal{K}=-1$ model) are of the order of 1 Mpc which is very close
to the observations \cite{Kar2012}. Outside of these spheres, the dragging effect of the gravitational attraction goes very fast to zero. We have also shown that the
presence of the cosmological constant is not absolutely crucial for the Hubble flows.
The Hubble flows may also take place in the absence of the cosmological constant. The reason for the Hubble flow is the global cosmological expansion of the Universe.
However, the cosmological constant provides the accelerating motion of the Hubble flows. Moreover, it makes sense to talk about the zero-acceleration surface only if at
present time $\Omega_{\Lambda}>\Omega_M/2$. Additionally, the cosmological constant also reduces smearing of the Hubble flow by peculiar velocities (see, e.g., the
numerical simulation in \cite{Chernin2004}). Therefore, the observations of the Hubble flows even at a few Mpc may reveal the presence of dark energy in the Universe
\cite{Sand2,Sand3}.

Now, we describe the differences between two our models. The main one consists in the spatial distribution of the inhomogeneities (gravitating sources). In the model
with $\mathcal{K}=0$, we were forced to introduce the radius $R_0$ around a gravitating mass to ensure the finiteness of the gravitational potential at infinity. This
radius is completely defined by the mass $m_0$ of the gravitating source and the cosmological parameters $H_0$ and $\Omega_M$ (at present time). Then, in such toy model,
all inhomogeneities are surrounded by these spheres which we call the local gravity ones because outside of them the gravitational potential is identically equal to zero
and we turn back to the unperturbed background Friedmann model with averaged ideal cosmological medium and the Hubble flows in it. In this model, the motion of a test
body inside of the local gravity sphere occurs in full analogy with the Schwarzschild-de Sitter model where the acceleration is completely defined by the cosmological
constant (without any admixture of dark matter); the zero-acceleration radius $R_*$ is at present time inside of the local gravity sphere: $R_*<\overline{R}_0$ if
$\Omega_{\Lambda}>\Omega_M/2$, and the dragging effect is absent for $R\ge R_0$. The spatial distribution of matter, where inhomogeneities are surrounded by empty
spheres of local gravity and ideal cosmological medium is outside of these spheres, looks a little bit artificial. Therefore, we have considered the hyperbolic model
($\mathcal{K}=-1$). Here, the inhomogeneities are distributed completely randomly. The finiteness of the gravitational potential for any number of gravitating sources
follows naturally from the solution of the Poisson equation without any artificial cutoff. Outside of the zero-acceleration surface (with the radius $\overline{R}_H$ at
present time), the dragging effect is not absent but goes asymptotically (rather fast) to zero. In contrast to the previous model, the gravitational interaction between
distant inhomogeneities is not absent, although it is suppressed by cosmological expansion. We have also shown that the superposition principle for the nonrelativistic
gravitational interaction works here.
Therefore, our model gives a possibility to simulate the dynamical behavior of an arbitrary number of randomly distributed inhomogeneities inside of the cell of
uniformity taking into account both the gravitational interaction between them and cosmological expansion of the Universe. Some simple examples of such simulations are
depicted on Figures \ref{K=-1}-\ref{distances}.
This is the main result of our paper.


\acknowledgments

This work was supported in part by the "Cosmomicrophysics-2" programme of the Physics and Astronomy Division of the National Academy of Sciences of Ukraine. We want to
thank the referee for his/her comments which have considerably improved the motivation of our investigations and the presentation of the results.


\end{document}